\documentclass [12pt] {article}
\usepackage{a4}
 
\usepackage{amsfonts}
\usepackage{amsmath} 
\usepackage{graphics}
\usepackage{epsfig}
\usepackage{epic}

\author{T.~Leonhardt, W.~R\"uhl}
\title{General graviton exchange graphs for four-point functions in
the AdS/CFT correspondence}

\begin{document}

\newcommand {\eps}{\varepsilon}
\newcommand {\ti}{\tilde}
\newcommand {\D}{\Delta}
\newcommand {\G}{\Gamma}
\newcommand {\de}{\delta}
\newcommand {\al}{\alpha}
\newcommand {\la}{\lambda}
\thispagestyle{empty}

\noindent hep-th/0211092   \hfill  November 2002 \\

\noindent
\vskip3.3cm
\begin{center}

{\Large\bf On the proposed AdS dual of the critical
$\boldsymbol{O(N)}$ sigma model for any dimension
$\boldsymbol{2\!<\!d\!<\!4}$ }
%\footnote{Supported in part by the `Deutsche Forschungsgemeinschaft'}}\\
\bigskip\bigskip\bigskip

{\large Thorsten Leonhardt, Ahmed Meziane and Werner R\"uhl}
\medskip

{\small\it Department of Physics\\
     Erwin Schr\"odinger Stra\ss e \\
     University of Kaiserslautern, Postfach 3049}\\
{\small\it 67653 Kaiserslautern, Germany}\\
\medskip
{\small\tt tleon,meziane,ruehl@physik.uni-kl.de}
\end{center}

\bigskip %\bigskip
\begin{center}
{\sc Abstract}
\end{center}
\noindent We evaluate the $4$-point function of the auxiliary field in the
critical $O(N)$ sigma model at $O(1/N)$ and show that it describes the
exchange of tensor currents of arbitrary even rank $l>0$. These are
dual to tensor gauge fields of the same rank in the AdS theory, which
supports the recent hypothesis of Klebanov and Polyakov. Their
couplings to two auxiliary fields are also derived.

\newpage

%%%%%%%%%%%%%%%%%%%%%%%%%%%%%%%%%%%%%%%%%%%%%%%%%%%%%%%%%%%%%%%%%%%%%%

\section{Introduction}
One of the outstanding problems in gauge theory is to understand
their low energy behaviour. Since these theories are
asymptotically free, standard perturbative techniques do not apply.

In the last five years we learned a lot about the strong coupling
behaviour of supersymmetric Yang Mills theory in the large $N$ limit,
where $N$ is the number of colors. This progress is based on
Maldacena's conjecture
\cite{Maldacena:1997re,Gubser:1998bc,Witten:1998qj}, which states that
four dimensional $\mathcal{N}\!=\!4$ supersymmetric Yang Mills theory
in the large $N$ and large 't Hooft coupling $\lambda$ limit is dual
to type IIB supergravity in the AdS$_5 \times S^5$ background. This
conjecture has been successfully tested and generalized in many ways
(see
\cite{Aharony:1999ti,D'Hoker:2002aw} and references therein). Now the next
logical step would be the consideration of the small $\lambda$ limit
of the conjecture. The theory describing this limit is a theory
containing fields of arbitrary high spins
\cite{Mikhailov:2002bp,Fradkin:ks,Fradkin:1986qy}. Unfortunately,
these theories are not very well understood for general dimensions.

Thus it is reasonable to study the AdS/CFT correspondence at simpler
models in order to find extensions of the duality on the
one hand and to obtain statements about higher spin theories on the other.

In a recent work by Klebanov and Polyakov \cite{Klebanov:2002ja} a
conjecture about the dual theory of the critical $O(N)$ sigma model
was formulated. The claim is that if one considers the four
dimensional minimal higher spin gauge theory with symmetry group $hs(4)$
in AdS$_4$ and applies the standard AdS/CFT procedure, then one
obtains correlation functions of some fields in the $O(N)$ sigma model
at its critical point, which is well known to be a conformal field
theory.

In this letter we want to add some evidence supporting this
conjecture. The minimal higher spin gauge theory consists of abelian
gauge fields $h^{(l)}$ which are symmetric tensor fields of even rank
$l\ge0$.  Among these, we denote the scalar field by
$\sigma:=h^{(0)}$.  On the other side of the duality there is the
critical $O(N)$ sigma model, which has as fundamental degrees of
freedom the Heisenberg spin field $\vec \phi$ and the auxiliary field
$\al$.

According to the general AdS/CFT procedure, the boundary values of
the gauge fields are sources for conformal field theory operators,
which are symmetric tensor fields of the same rank as the
source. In the model investigated by Klebanov and Polyakov the scalar
field in AdS has the operator $\vec \phi \cdot \vec \phi$ as boundary
value and is thus dual to the auxiliary field $\al$ in the sigma
model. The other higher spin fields $h^{(l)}$ in AdS with $l\ge2$
are dual to some spacetime tensor operators $J^{(l)}$, which are
bilinears in the $\phi$-field, such that they transform as singlets
under $O(N)$. Moreover, the conformal dimensions are fixed by the
requirement that the gauge fields are massless:
\begin{align}\label{dimmasslgauge}
\D_{J^{(l)}}= d +l-2.
\end{align}
By inspecting the $\al$ four-point function up to order $O(1/N)$ obtained from a sigma
model computation, we find that the only exchanged fields, which are neither
$\alpha$ nor $\al$ composites, are the duals of the higher spin gauge
fields. We emphazise that this is far from being trivial, since in
general one would expect all fields with conformal dimensions
\begin{align}
\D =  d +l-2+2t,\; t \in \mathbb{N}
\end{align}
to contribute to the four-point function. Moreover we have been able
to compute all couplings $\gamma_l$ of two $\alpha$ fields to an
exchanged tensor $J^{(l)}$. Now it follows by the
duality that all three point couplings of the gauge fields in AdS are
fixed, which is compatible with gauge invariance \cite{Vasiliev:1995dn}. 

Furthermore, we were able to solve the puzzle of the shadow field for
this model by showing that the shadow field of the $\al$ field, which
inevitably appears in $\al$ exchanges, is part of the amplitudes,
where the composite tensor operators $J^{(l)}$ are exchanged.

Our results rely heavily on the detailed knowledge about the critical
$O(N)$ sigma model gained in \cite{Lang:1992pp,LR} (see also the
references therein).

%%%%%%%%%%%%%%%%%%%%%%%%%%%%%%%%%%%%%%%%%%%%%%%%%%%%%%%%%%%%%%%%%%%%%%

\section{The critical sigma model}

We formalate the conformal field theory corresponding to the critical
sigma model in $d$ spacetime dimensions with $2 \! < \! d \! < \! 4$ by two bosonic
fundamental fields: the scalar Heisenberg spin field $\vec \phi$,
which transforms as an $O(N)$ vector, and the scalar auxiliary field
$\alpha$, which is a singlet under $O(N)$. Their propagators are
given by
\begin{align}\label{props}
\langle \phi_a(x_1) \phi_b(x_2) \rangle & = A \delta_{ab}\bigl( x_{12}^2
\bigr)^{-\de}, \;\textrm{denoted by a solid line} \nonumber \\
\langle \alpha(x_1) \alpha(x_2) \rangle & = B \bigl( x_{12}^2 \bigr)^{-\beta},
\;\textrm{denoted by a dashed line},
\end{align}
where $\de$ and $\beta$ are the conformal dimensions of the
respective fields, $A$ and $B$ are normalization constants and $a,b$
are indices of the vector representation of $O(N)$. These fields
interact via
\begin{align}
z^{\frac{1}{2}}  \alpha(x) \vec \phi(x) \cdot \vec \phi(x).
\end{align}
The conformal dimensions are composed of a canonical part and an
anomalous contribution, which can be expanded as a series in
$1/N$:
\begin{align}
\delta & = \mu -1 +\eta , \qquad\quad \eta =  \sum_{k=1}^\infty
\frac{\eta_k}{N^k} \nonumber \\
\beta & = 2 -2 \eta -2 \kappa, \qquad \kappa =  \sum_{k=1}^\infty
\frac{\kappa_k}{N^k},
\end{align}
where we introduced the convenient abbreviation
\begin{align}
\mu = \frac{d}{2}.
\end{align}
If $A$ and $B$ in (\ref{props}) are absorbed in the coupling constant
(which is assumed from now on), the normalizations of the propagators
equal one and the coupling constant takes a fixed value at the
critical point. This fixed value can also be expanded in $1/N$:
\begin{align}
z= \sum_{k=1}^\infty \frac{z_k}{N^k}.
\end{align}
The coefficients $\eta_k, \kappa_k, z_k$ are functions of $d$. Among
them, $\eta_1, \eta_2, \eta_3, \kappa_1, z_1, z_2$ have been computed
\cite{Fradkin:1990dk,Vasilev:yc,Vasilev:dg,Vasilev:dc}. In the present
context we only need  
\begin{align}
z_1 = 2 \pi^{-2\mu} \frac{(\mu-2) \G(2\mu-2)}{\G(\mu)\G(1-\mu)}
\end{align}

With these propagators and this interaction, the four-point function
of the $\alpha$ fields can be computed perturbatively. At order $1/N$
there are nine graphs which contribute to the four-point function:

\noindent At order $1$, we have
\begin{align}
\begin{picture}(60,40)
\put(20,-20){
\dashline{4}(-30,50)(20,50)
\dashline{4}(-30,0)(20,0)
\put(-40,46){$1$}
\put(25,46){$2$}
\put(-40,-4){$3$}
\put(25,-4){$4$}
\put(-10,-20){$A_1$}}
\end{picture}
\qquad\qquad
\begin{picture}(60,40)
\put(20,-20){
\dashline{4}(-30,50)(-30,0)
\dashline{4}(20,50)(20,0)
\put(-40,46){$1$}
\put(25,46){$2$}
\put(-40,-4){$3$}
\put(25,-4){$4$}
\put(-10,-20){$A_2$}}
\end{picture}
\qquad\qquad
\begin{picture}(60,40)
\put(20,-20){
\dashline{4}(-30,50)(20,0)
\dashline{4}(-30,0)(-11,19)
\dashline{4}(-1,29)(20,50)
\put(-40,46){$1$}
\put(25,46){$2$}
\put(-40,-4){$3$}
\put(25,-4){$4$}
\put(-10,-20){$A_2$}}
\end{picture}
\end{align}
\vspace{8mm}

\noindent At order $O(1/N)$ there are the exchange graphs
\vspace{5mm}
\begin{align}
\begin{picture}(60,30)
\put(-10,10){
\dashline{4}(0,30)(10,20)
\dashline{4}(30,20)(40,30)
\put(10,20){\line(1,0){20}}
\put(10,20){\line(1,-1){10}}
\put(30,20){\line(-1,-1){10}}
\dashline{4}(20,10)(20,-10)
\put(20,-10){\line(-1,-1){10}}
\put(20,-10){\line(1,-1){10}}
\put(10,-20){\line(1,0){20}}
\dashline{4}(10,-20)(0,-30)
\dashline{4}(30,-20)(40,-30)
\put(-10,-35){$3$}
\put(-10,28){$1$}
\put(45,-35){$4$}
\put(45,28){$2$}
\put(15,-40){$B_{11}$}}
\end{picture}
\qquad \qquad\!\!\!
\begin{picture}(60,30)
\put(-0,0){
\dashline{4}(-10,30)(0,20)
\dashline{4}(-10,-10)(0,0)
\line(0,1){20}
\put(0,0){\line(1,1){10}}
\put(0,20){\line(1,-1){10}}
\dashline{4}(10,10)(30,10)
\put(30,10){\line(1,1){10}}
\put(30,10){\line(1,-1){10}}
\put(40,0){\line(0,1){20}}
\dashline{4}(40,0)(50,-10)
\dashline{4}(40,20)(50,30)
\put(-19,-15){$3$}
\put(-19,28){$1$}
\put(53,-15){$4$}
\put(53,28){$2$}
\put(15,-30){$B_{12}$}}
\end{picture}
\qquad\qquad
\begin{picture}(60,30)
\put(-0,10){
\dashline{4}(-10,30)(10,-20)
\dashline{4}(30,20)(40,30)
\put(10,20){\line(1,0){20}}
\put(10,20){\line(1,-1){10}}
\put(30,20){\line(-1,-1){10}}
\dashline{4}(20,10)(20,-10)
\put(20,-10){\line(-1,-1){10}}
\put(20,-10){\line(1,-1){10}}
\put(10,-20){\line(1,0){20}}
\dashline{4}(10,20)(5,8)
\dashline{4}(0,-5)(-10,-30)
\dashline{4}(30,-20)(40,-30)
\put(-20,-35){$3$}
\put(-20,28){$1$}
\put(45,-35){$4$}
\put(45,28){$2$}
\put(10,-40){$B_{13}$}}
\end{picture}
\end{align}
\vspace{5mm}

\noindent and the box graphs

\begin{align}
\begin{picture}(80,30)
\put(10,10){
\dashline{4}(0,30)(10,20)
\dashline{4}(50,30)(40,20)
\put(10,20){\line(1,0){30}}
\put(10,20){\line(0,-1){30}}
\dashline{4}(0,-20)(10,-10)
\dashline{4}(50,-20)(40,-10)
\put(40,-10){\line(-1,0){30}}
\put(40,-10){\line(0,1){30}}
\put(-8,-25){$3$}
\put(-8,28){$1$}
\put(53,-25){$4$}
\put(53,28){$2$}
\put(17,-40){$B_{21}$}}
\end{picture}
\qquad  \quad
\begin{picture}(80,30)
\put(5,10){
\dashline{4}(0,30)(10,20)
\dashline{4}(60,30)(40,-10)
\put(10,20){\line(1,0){30}}
\put(10,20){\line(0,-1){30}}
\dashline{4}(0,-20)(10,-10)
\dashline{4}(60,-20)(50,-1)
\dashline{4}(45,9)(40,20)
\put(40,-10){\line(-1,0){30}}
\put(40,-10){\line(0,1){30}}
\put(-8,-25){$3$}
\put(-8,28){$1$}
\put(63,-25){$4$}
\put(63,28){$2$}
\put(17,-40){$B_{22}$}}
\end{picture}
\qquad \quad
\begin{picture}(80,30)
\put(-5,15){
\dashline{4}(0,30)(10,20)
\dashline{4}(50,30)(40,20)
\put(10,20){\line(1,0){30}}
\put(10,20){\line(0,-1){30}}
\dashline{4}(0,-30)(40,-10)
\dashline{4}(20,-15)(10,-10)
\dashline{4}(50,-30)(32,-21)
\put(40,-10){\line(-1,0){30}}
\put(40,-10){\line(0,1){30}}
\put(-8,-35){$3$}
\put(-8,28){$1$}
\put(53,-35){$4$}
\put(53,28){$2$}
\put(17,-45){$B_{23}$}}
\end{picture}
\end{align}
\vspace{5mm}

\noindent Then we obtain for the $\al$ four point function including
terms up to order $O(1/N)$

\begin{multline}\label{alpha4pt}
\langle \alpha(x_1) \alpha(x_2) \alpha(x_3) \alpha(x_4) \rangle = A_1
+ A_2 + A_3 \\
+ \frac{1}{N} \biggl\{ z_1^3 \Bigl(B_{11}+B_{12}+B_{13}\Bigr)
+ z_1^2\Bigl(B_{21}+B_{22}+B_{23}\Bigr) \biggr\} + O(\frac{1}{N^2}).
\end{multline}
The $A$ graphs consist only of two-point functions, therefore they are
of order 1, and in the $B$ graphs we have six (resp. four) vertices, each
giving a factor of $N^{-1/2}$, and two (resp. one) closed $\phi$
loops, each contributing a factor of $N$. Thus the $B$ graphs are
indeed of order $1/N$. These graphs are evaluated in terms of the
conformally invariant variables
\begin{align}
u=\frac{x_{13}^2 x_{24}^2}{x_{12}^2 x_{34}^2}, \quad v=\frac{x_{14}^2
x_{23}^2} {x_{12}^2 x_{34}^2}.
\end{align}

Next we will apply an operator product expansion in the channel
\begin{align}
x_{13}=x_1-x_3 \rightarrow 0, \quad x_{24}=x_2-x_4 \rightarrow 0,
\end{align}
which is the same as 
\begin{align}\label{schannel}
u \rightarrow 0, \quad v \rightarrow 1.
\end{align}
Operator product expansions in other channels correspond to analytic
continuations of the above result and present no further problem.

In the OPE of (\ref{alpha4pt}) in the above channel, composite fields
with approximate (exact at $N=\infty$) dimensions $\D^{(0)}$, which
are symmetric traceless tensors of even rank $l$ show up. These are
operators composed of two $\al$ fields with approximate dimensions 
\begin{align}\label{alphacomp}
\D_{2\al}^{(0)} = 4 +2t +l,
\end{align}
and operators composed of two $\phi$ fields with approximate dimensions
\begin{align}\label{phicomp}
\D_{2\phi}^{(0)} = 2\mu -2  +2t +l.
\end{align}
These composite operators can be represented as towers. We denote the
tower corresponding to the composite operators of $\phi$ fields as
tower I and the one of the $\al$ fields as tower II. In the OPE there
are furthermore the unit operator (graph $A_2$) and the $\al$ field
itself (graph $B_{12}$) exchanged. The anomalous part of the
dimensions of the composite fields from tower II are of order $O(1/N)$
and can be read off from the coefficients of the $log \;u$ terms in
the $B$ graphs. The anomalous dimensions of the composite operators in
tower I can only be obtained from a calculation including terms of
order $O(1/N^2)$.

One of the main results of the present investigation is the
observation that only those composite fields from tower I
(\ref{phicomp}) with 
\begin{align}
t=0, \quad l>0 \textrm{ and even}
\end{align}
appear in the OPE. 
Therefore we only have tensor fields with conformal dimensions 
\begin{align}\label{dimt0}
\D^{(0)}_{2\phi}\rvert_{t=0}=2\mu -2 +l,
\end{align}
to first order $O(1/N)$.  By comparison with (\ref{dimmasslgauge}) we
observe that the dimensions (\ref{dimt0}) agree with those of the
fields dual to the gauge fields in AdS. Thus the tensors in the tower
(\ref{phicomp}) with $t=0$ are to be identified with the duals of the
gauge fields.

If these were the exact dimensions of tensor fields from tower I with
$t=0$, then these fields were conserved currents. However, at this
stage we have no argument for the vanishing of their anomalous
dimensions, except for the tensor of rank two. On the CFT side it
corresponds to the energy momentum tensor, which is conserved and has
no anomalous dimension. On the AdS side this field corresponds to the
fluctuations of the metric around the AdS background
\begin{align}
g_{\mu \nu} = g_{\mu \nu}^{AdS} + h^{(2)}_{\mu \nu},
\end{align}
which must be massless due to the masslessness of the gravitons. By
definition of the energy momentum tensor $T_{\mu \nu}=J^{(2)}_{\mu
\nu}$ it follows that $h^{(2)}$ and $J^{(2)}$ are indeed dual.

Now we turn to the calculation of the coupling constants of two $\al$
fields to the tensors $J^{(l)}$, which will also
imply the result that only tensors of tower I with $t=0$ appear in
the OPE of the $\al$ four-point function. Most results of
the actual calculation are taken from \cite{Lang:1992pp}, and we use the notation
of this reference. 

The $\al$ four-point function can be written as a sum of exchange
graphs in the direct and crossed channels. The exchanged fields are
the $\al$ and the composite fields of both towers I and II. The
contributions of tower (\ref{phicomp}) come from the $c_{nm}^{(12)},
c_{nm}^{(21)}$ and $c_{nm}^{(22)}$ terms of the respective graphs
$B_{12}, B_{21}$ and $B_{22}$:
\begin{align}
(x_{12}^2 x_{34}^2)^{-\beta} u^{\de-\beta} \sum_{n,m \ge 0}
\frac{u^n(1-v)^m}{n!\,m!} c_{nm}^{(ij)},
\end{align}
together with a constant factor split off to achieve the normalization
$c_{00}^{(ij)}=1$. Then we get for the contributions of the tensors in
the tower I, which are the composites of the $\vec \phi$ fields, to
the $\al$ four-point function
\begin{multline}\label{alpha4poit}
\langle \al(x_1) \cdots \al(x_4) \rangle\Bigr\rvert_{I} = \frac{1}{N}
\Bigl(z_1^3 B_{12} + z_1^2 \bigl(B_{21} +B_{22}\bigr)
\Bigr)\Bigr\rvert_{I} \\
= (x_{12}^2 x_{34}^2)^{-\beta} u^{\de-\beta} \sum_{n,m \ge 0}
\frac{u^n(1-v)^m}{n!\,m!} 4(2 \mu-3)^2 \Bigl(-2 c_{nm}^{(12)}
+ c_{nm}^{(21)} + c_{nm}^{(22)} \Bigr),
\end{multline}
where \footnote{ In the equation for $c_{nm}^{(22)}$ a misprint in
\cite{Lang:1992pp}, namely a wrong factor of $\frac{1}{n!}$ in eq. (C.10) was corrected.}
\begin{align}\label{alpha4ptco}
c_{nm}^{(12)} & = \frac{(\mu-1)_n [(\mu-1)_{n+m}]^2} {(2
\mu-2)_{2n+m}}, \nonumber \\
c_{nm}^{(21)} & = \frac{n!\, [(\mu-1)_{n+m}]^2} {(2 \mu-3)_{2n+m}}
\sum_{s=0}^n \frac{(2 \mu-4)_{n+m+s}(\mu-2)_s}{s!\, (2\mu-3)_{n+m+s}},
\nonumber \\
c_{nm}^{(22)} & = \frac{(n+m)!\, (\mu-1)_n (\mu-1)_{n+m}} {(2 \mu-3)_{2n+m}}
\sum_{s=0}^{n+m} \frac{(2 \mu-4)_{n+s}(\mu-2)_s}{s!\, (2\mu-3)_{n+s}}. 
\end{align}
If we denote 
\begin{align}
C_{nm} = -2 c_{nm}^{(12)} + c_{nm}^{(21)} + c_{nm}^{(22)}
\end{align}
then it is easy to see that $C_{00}=C_{01}=0$, implying that the
contributions of tower I to the $\al$ four-point function at order
$O(1/N)$ starts with a tensor of rank $2$. We note that the $1/N$
dependence is hidden in the normalization of the $c^{(ij)}_{nm}$.

On the other hand, with the help of the `master formula' of \cite{LR},
eqns. (2.16)-(2.20), we can compute the four-point exchange amplitude
$W_{\beta}(x_1,\ldots,x_4;\D,l)$ of a rank $l$ symmetric tensor field of
dimension $\D$ from (\ref{dimt0}) with four $\al$ fields of dimension
$\beta$ as external legs. With the ad hoc normalization
\begin{align}
\al_{0l}^{(l)}=1
\end{align}
we get
\begin{align}\label{tensex}
W_{\beta}(x_1,\ldots,x_4;\D,l) = (x_{12}^2 x_{34}^2)^{-\beta} u^{\de-\beta}
\sum_{n,m \ge 0} \frac{u^n(1-v)^m}{n!\,m!} \al_{nm}^{(l)}
\end{align}
with
\begin{align}\label{tensexco}
\al_{nm}^{(l)} = \sum_{s=0}^n (-1)^s \binom{n}{s} \binom{m+n+s}{l}
\frac{(\mu-1+l)_{m-l+n}(\mu-1+l)_{m-l+n+s}}{(2\mu-2+2l)_{m-l+n+s}},
\end{align}
where we define the Pochhammer symbols by
\begin{align}
(z)_n = \frac{\G(z+n)}{\G(z)}\quad \textrm{for all}\; n \in \mathbb{Z}.
\end{align}
Now we look for the coupling constants of the $\al$ fields to the rank
$l$ tensor fields with the dimension $\D$ from (\ref{dimt0}). We do
this by making the ansatz
\begin{align}\label{anseqtosolve}  
\langle \al(x_1) \cdots \al(x_4) \rangle\Bigr\rvert_{I} =
\sum_{\substack{l>0 \\ l \,\textrm{even}}} \gamma_l^2
W_{\beta}(x_1,\ldots,x_4;\D,l)
\end{align}
which reduces after insertion of eqns (\ref{alpha4poit}),
(\ref{alpha4ptco}), (\ref{tensex}) and (\ref{tensexco}) to 
\begin{align}\label{eqtosolve}
4 (2 \mu-3)^2 C_{nm} = \sum_{l>0} \gamma_l^2 \al_{nm}^{(l)}.
\end{align}
It looks like a surprise that this equation has a solution at all,
because from the CFT point of view one expects the necessity of
further exchange amplitudes of tensors with dimensions (\ref{phicomp})
with $t>0$ on the right hand side. But we indeed solved this equation
via computer algebra up to tensor rank 8 and obtained the solution
($l\ge2$ and even)
\begin{align}\label{soleqtosolve}
\gamma_l^2 = \frac{(l!)^2} {2^{3l-5}} \frac{(\mu
+\frac{l}{2}-1)_{\frac{l}{2}}} {(\mu-\frac{1}{2})_{l-1} (\mu
-\frac{1}{2})_{\frac{l}{2}-1}}.
\end{align}
Now that we know that a solution to this equation exists, we can also
obtain it by setting $n=0$ in (\ref{eqtosolve}) and inverting the
triangular matrix $\al_{0m}^{(l)}$:
\begin{align}
\sum_{l=0}^m \sigma_l^{(k)} \al_{0m}^{(l)} = \de_m^k.
\end{align}
The solution to this equation is given by
\begin{align}
\sigma_l^{(k)} = (-1)^{l-k} \binom{l}{k}
\frac{[(\mu-1+k)_{l-k}]^2} {(2\mu-3+k+l)_{l-k}}.
\end{align}
Thus we also have an algebraic check for the extrapolated results from
the computer algebra.

Let us finally mention that all these results are based on a $1/N$
expansion, and that we only considered the first order $O(1/N)$. We
expect these structures to be modified in the next perturbative
order. In particular, there are contributions of the tower
(\ref{phicomp}) with $t>0$ to the $\al$ four-point function, e.g. from
a graph like
\begin{align}
\begin{picture}(300,34)
\put(-40,-10){\put(174,4){\framebox(30,30){}}
\put(-20,0){\dashline{4}(160,4)(190,4)}
\put(-20,30){\dashline{4}(160,4)(190,4)}
\put(48,0){\dashline{4}(160,4)(190,4)}
\put(48,30){\dashline{4}(160,4)(190,4)}
\put(16,16){\dashline{4}(160,4)(186,4)}
%\put(140,4){\circle*{3}}
%\put(140,34){\circle*{3}}
%\put(238,4){\circle*{3}}
%\put(238,34){\circle*{3}}
}
\end{picture}
\end{align}
\noindent at order $O(1/N^2)$.

%%%%%%%%%%%%%%%%%%%%%%%%%%%%%%%%%%%%%%%%%%%%%%%%%%%%%%%%%%%%%%%%%%%%

\section{Consequences for the AdS dual}

According to the proposal of Klebanov and Polyakov the critical $O(N)$
sigma model is the AdS dual of the minimal higher spin theory with
symmetry group $hs(4)$. This is a theory of infinitely many abelian gauge
fields, which are symmetric tensor fields of rank $l$ with
$l=0,2,4,\ldots$, each of these ranks occurring once. Due to gauge
invariance, these fields are supposed to be massless, at least at
first order in a perturbative expansion. Although the nonlinear action
is known, we have no information about the concrete form and values of
the couplings at hand. Nevertheless, the cubic couplings are supposed
to be fixed by gauge invariance.

The results of the previous section fit nicely into the scheme of
Klebanov's and Polyakov's proposal, because to first order the only
exchanged fields, which are not composite operators of $\al$ fields,
are those which are dual to the massless gauge fields in
AdS. Furthermore, any tensor field in the OPE of the $\al$ four-point
function with $t>0$ would imply that more than one field of a given
tensor rank, which is not a composite operator of $\al$ fields, must
exist in the AdS theory. But this would be incompatible with the
higher spin theory, since any tensor rank may appear at most once.

We assume the interactions to be local and write down Witten graphs to
reproduce the behaviour of the $\al$ four-point function. To this end
we first consider the graph $B_{12}$ in the sigma model, which is
one-particle reducible with respect to the $\al$ field in the channel
(\ref{schannel})
\begin{align}
B_{12} \sim \sum_{n,m \ge 0} \frac{u^n(1-v)^m}{n!\,m!}
\Bigl(u^{\de-\beta} c_{nm}^{(12)} + u^{\mu-\de-\beta} d_{nm}\Bigr).
\end{align} 
Since $\al$ is a fundamental field, the $\al$
exchange graph must contain besides the ``direct'' term, which is the
series with the coefficients $d_{nm}$, also the ``shadow'' term,
corresponding to the coefficients $c_{nm}^{(12)}$. We can reproduce the
singular behavior of the CFT direct channel by a Witten graph 
\begin{align}
\begin{picture}(40,50)
\put(20,20){{\oval(100,50)}}
\dashline{4}(-24,1)(0,20)
\dashline{4}(-24,39)(0,20)
\dashline{4}(0,20)(40,20)
\dashline{4}(40,20)(63,0)
\dashline{4}(40,20)(63,39)
\end{picture}
\end{align}
if we introduce a cubic interaction in AdS
\begin{align}
g_3 \sigma^3, \quad g_3 = O(N^{-1/2}).
\end{align}
This graph produces the correct direct channel, but the shadow term is
missing. To solve this problem, we note that since the tensor fields
$J^{(l)}$ are dual to the gauge fields, the sum of the exchange diagrams of the
tensor fields $J^{(l)}$ are to be recovered in the following sum of
Witten graphs
\begin{align}\label{sumofwitgr}
\sum_{\substack{l>0\\l \textrm{even}}}\qquad
\begin{picture}(100,30)
\put(15,-20){\put(20,20){\oval(100,50)}
\dashline{4}(-24,1)(0,20)
\dashline{4}(-24,39)(0,20)
\dottedline{3}(0,20)(40,20)
\dashline{4}(40,20)(63,0)
\dashline{4}(40,20)(63,39)
\put(14,22){$h^{(l)}$}}
\end{picture},
\end{align}
%\vspace{3mm}

\noindent
with the couplings
\begin{align}
f_{2,l} \sigma^2 h^{(l)}, \quad f_{2,l} = O(N^{-1/2}).
\end{align}
In the previous section we have seen that the shadow term of the
$\alpha$ field, which corresponds to the coefficients $c_{nm}^{(12)}$,
is part of the coefficients $C_{nm}$, which is related by the solution
of eq. (\ref{eqtosolve}) to the sum of the exchange graphs of the tensors
$J^{(l)}$. Thus the shadow term of the $\al$ exchange is loosely
speaking `hidden in the tensor exchange', and the same applies to the
boundary theory of the higher spin theory in AdS. The fact that the
shadow term of the $\al$ exchange graph is part of the tensor exchange
graphs can be explained in the following way: the graphs $B_{21}$ and
$B_{22}$ with the coefficients $c_{nm}^{(21)}$ and $c_{nm}^{(22)}$
contain one $\phi$-loop. The shadow field of the $\al$ field is $\vec
\phi \cdot \vec \phi$ in the sigma model, whereas the rank $l$ tensors
$J^{(l)}$ arise by application of differential operators of order $l$
on each factor of $\vec \phi \cdot \vec \phi$ (at leading order). Thus
the shadow term fills up the $l=0$ gap in the tower of conserved
currents in a natural way.

Now we know that the couplings of the tensors $J^{(l)}$ to the $\al$
field in the CFT are fixed by (\ref{soleqtosolve}) and the cubic $\al$
coupling is fixed by the above argument based on shadow
symmetry. Thus, if Klebanov's and Polyakov's conjecture is true, then
we have a further argument for all cubic couplings of the gauge fields
to be fixed, which is compatible with gauge invariance. 

Analogous considerations apply to $(n_1+n_2)$-point functions of $\al$
fields with an OPE connecting a set of $n_1$ with a set of $n_2$ $\al$
fields
\begin{align}
\sum_{\Psi} \qquad \quad\qquad \quad\quad
\begin{picture}(40,30)
\put(-40,-20){
\put(0,20){\circle*{10}}
\dashline{4}(-24,1)(0,20)
\dashline{4}(-26,6)(0,20)
\dashline{4}(-28,11)(0,20)
\put(-22,20){\circle*{2}}
\put(-22,25){\circle*{2}}
\put(-22,30){\circle*{2}}
\dashline{4}(-24,39)(0,20)
\put(0,20){\line(1,0){40}}
\put(40,20){\circle*{10}}
\dashline{4}(40,20)(63,0)
\dashline{4}(68,5)(40,20)
\dashline{4}(70,11)(40,20)
\put(62,20){\circle*{2}}
\put(62,25){\circle*{2}}
\put(62,30){\circle*{2}}
\dashline{4}(40,20)(63,39)
\put(15,22){$\Psi$}}
\end{picture},
\end{align}

\vspace{3mm}
\noindent where there are $n_1$ $\al$-legs on the left and $n_2$ on
the right. The connected $(n_1+n_2)$-point function with an
$\al$-exchange is given in lowest order by a graph with two
$\phi$-loops
\begin{align}
\begin{picture}(100,30)
\put(40,-20){
\put(0,20){\circle{30}}
\dashline{4}(-30,0)(-15,10)
\dashline{4}(-32,5)(-17,12)
\dashline{4}(-34,10)(-19,15)
\put(-28,20){\circle*{2}}
\put(-28,25){\circle*{2}}
\put(-28,30){\circle*{2}}
\dashline{4}(-30,39)(-15,30)
\dashline{4}(16,20)(64,20)
\put(80,20){\circle{30}}
\dashline{4}(94,10)(108,0)
\dashline{4}(96,12)(110,5)
\dashline{4}(98,15)(112,10)
\put(107,20){\circle*{2}}
\put(107,25){\circle*{2}}
\put(107,30){\circle*{2}}
\dashline{4}(94,30)(108,39)
\put(36,22){$\al$}
}\end{picture}\qquad\qquad\qquad.
\end{align}
\vspace{3mm}

\noindent Again, the direct term is produced by a Witten graph like e.g.
\begin{align}
\begin{picture}(120,30)
\put(40,-20){\put(20,20){\oval(100,50)}
\dashline{4}(-24,1)(0,20)
\dashline{4}(-29,10)(-16,8)
\dashline{4}(-29,17)(-9,13)
\dashline{4}(-29,25)(-14,31)
\dashline{4}(-24,39)(0,20)
\dashline{4}(0,20)(40,20)
\dashline{4}(40,20)(63,0)
\dashline{4}(55,7)(69,11)
\dashline{4}(48,26)(69,20)
\dashline{4}(40,20)(63,39)
\put(17,22){$\sigma$}}
\end{picture}.
\end{align}
\vspace{3mm}

\noindent This graph is of the order $O(N^{-\frac{1}{2}(n_1+n_2-2)})$,
but if we replace some of the cubic
couplings by couplings like 
\begin{align}
g_n \sigma^n,\quad \textrm{with} \;g_n=O(N^{-\frac{1}{2}(n-2)})
\end{align}
the order in $1/N$ is preserved.
As before, there are $l$-rank tensor exchange amplitudes in CFT
absorbing the shadow term of the $\al$-exchange amplitude. These are
produced by graphs of the following type
\begin{align}
\begin{picture}(60,30)
\put(40,-20){
\put(0,20){\circle{30}}
\dashline{4}(-30,0)(-15,10)
\dashline{4}(-32,5)(-17,12)
\dashline{4}(-34,10)(-19,15)
\put(-28,20){\circle*{2}}
\put(-28,25){\circle*{2}}
\put(-28,30){\circle*{2}}
\dashline{4}(-30,39)(-15,30)
\put(-80,0){\dashline{4}(94,10)(108,0)
\dashline{4}(96,12)(110,5)
\dashline{4}(98,15)(112,10)
\put(107,20){\circle*{2}}
\put(107,25){\circle*{2}}
\put(107,30){\circle*{2}}
\dashline{4}(94,30)(108,39)}
}\end{picture}\qquad.
\end{align}
\vspace{3mm}

\noindent The Witten graphs which give these amplitudes are of the following form
\begin{align}
\sum_{\substack{l>0\\l \textrm{even}}}\qquad 
\begin{picture}(100,30)
\put(15,-20){\put(20,20){\oval(100,50)}
\dashline{4}(-24,1)(0,20)
\dashline{4}(-29,10)(-16,8)
\dashline{4}(-29,17)(-9,13)
\dashline{4}(-29,25)(-14,31)
\dashline{4}(-24,39)(0,20)
\dottedline{3}(0,20)(40,20)
\dashline{4}(40,20)(63,0)
\dashline{4}(55,7)(69,11)
\dashline{4}(48,26)(69,20)
\dashline{4}(40,20)(63,39)
\put(14,22){$h^{(l)}$}}
\end{picture},
\end{align}
\vspace{3mm}

\noindent
which are again of the order $O(N^{\frac{1}{2}(n_1+n_2-2)})$. If couplings like
\begin{align}
f_{n,l}\, \sigma^n h^{(l)}, \quad \textrm{with} \; f_{n,l} =
O(N^{-\frac{1}{2}(n-1)})
\end{align}
are used, the order $O(N^{\frac{1}{2}(n_1+n_2-2)})$ is maintained. 

However, shadow symmetry imposes certain polynomial relations for
every $n_1$ and $n_2$ on the coupling constants $f_{n,l}$ and $g_l$.

Finally, we want to fix the overall normalization of the Witten
four-point graphs corresponding to an exchange of a gauge field of
tensor rank $l$ (single summand of (\ref{sumofwitgr})). This can be
done in the following way. Although the action is not explicitly
given for our purpose, we make the assumption that the coupling of
two scalars to the graviton in the bulk is the usual one that comes
from the covariant derivatives in the kinetic terms of the scalar
fields. The resulting graviton exchange graph was calculated in
\cite{Leonhardt:2002ta} as a sum of power series in $(u,1-v)$ and we
display only that part, which contains terms singular in $u$ 
\begin{align}
G_{\textrm{sing.}}&(x_1, \ldots,x_4) = -\pi^{\mu} \tilde c K_{\D}^4 (x_{12}^2 x_{34}^2)^{-\D}
\Bigl(\frac{\D \G(\D+1-\mu)}{\G(\D)} \Bigr)^2 \sum_{n\ge0} \frac{u^{n+\mu-\D-1}}{n!} \nonumber \\
& \frac{\G( \mu-1+n)^3}{\G(2\mu+2n-2)}
\biggl\{\frac{1-\mu}{2\mu-1} F\Bigl[ {\mu-1+n, \mu-1+n \atop
2\mu+2n-2};\; 1-v \Bigr] \nonumber \\
& \qquad \qquad \qquad \qquad \quad +\frac{\mu+n-1}{2 \mu+2n-1} F\Bigl[ {\mu-1+n,
\mu+n \atop 2\mu+2n};\; 1-v \Bigr] \biggr\},  
\end{align}
where $\D$ denotes the conformal dimension of the external legs,
i.e. in our case $\D\!=\!\beta\!=\!2$, $\tilde c$ is the normalization
of the graviton propagator and $K_{\D}$ is the normalization of the
bulk-to-boundary propagator. As noted in
\cite{Leonhardt:2002ta}, this term exactly agrees with the exchange of
the energy momentum tensor in the boundary CFT. Thus we can compare
the coupling of the energy momentum tensor to the scalars in the
boundary CFT with that one from the sigma model by forming the
quotient of the $(n,m)=(0,2)$ summands:
\begin{align}\label{rel}
\frac{\langle \al \al \al \al \rangle}{G_{\textrm{sing.}}}\Bigl
\rvert_{(n,m)=(0,2)} = \frac{1}{N} \frac{2 \mu-1}{\pi^\mu \tilde c
K_2^4} \frac{\G(2 \mu)}{\G(\mu)^3 \G(3-\mu)^2}.
\end{align}
Note that there is a factor proportional to $1/N$ in the numerator of
the left hand side of (\ref{rel}), which comes from the natural
normalization of the $\al$ four point function.

By this procedure, together with the couplings of the tensors to the
scalars in the sigma model (\ref{soleqtosolve}), we fixed all
cubic couplings of the higher spin fields $h^{(l)}$ to the scalars
$\sigma$ in the boundary CFT, and thus also in the AdS theory.

\section{Conclusion and outlook}
As we have seen above, we have found a method of testing the
conjecture of Klebanov and Polyakov. Unfortunately, it cannot be
evaluated, since we have no explicit form of the interaction terms in
the AdS theory at hand. Thus it would be highly desireable to obtain
explicit expressions for these.

Our results are based on the inspection of the $\al$ four-point
function. However, we are also able to compute every $n$-point function of
the tensor fields $J^{(l)}$ at leading order $O(1/N)$.

\section*{Acknowledgement}
 
A.~M. would like to thank the German Academic Exchange Service (DAAD)
for financial support.

%%%%%%%%%%%%%%%%%%%%%%%%%%%%%%%%%%%%%%%%%%%%%%%%%%%%%%%%%%%%%%%%%%%%%%%


\begin{thebibliography}{99}

%\cite{Maldacena:1997re}
\bibitem{Maldacena:1997re}
J.~M.~Maldacena,
``The large $N$ limit of superconformal field theories and supergravity,''
Adv.\ Theor.\ Math.\ Phys.\  {\bf 2} (1998) 231
[Int.\ J.\ Theor.\ Phys.\  {\bf 38} (1999) 1113]
[arXiv:hep-th/9711200].
%%CITATION = HEP-TH 9711200;%% 

%\cite{Gubser:1998bc}
\bibitem{Gubser:1998bc}
S.~S.~Gubser, I.~R.~Klebanov and A.~M.~Polyakov,
``Gauge theory correlators from non-critical string theory,''
Phys.\ Lett.\ B {\bf 428} (1998) 105
[arXiv:hep-th/9802109].
%%CITATION = HEP-TH 9802109;%%

%\cite{Witten:1998qj}
\bibitem{Witten:1998qj}
E.~Witten,
``Anti-de Sitter space and holography,''
Adv.\ Theor.\ Math.\ Phys.\  {\bf 2} (1998) 253
[arXiv:hep-th/9802150].
%%CITATION = HEP-TH 9802150;%%

%\cite{Aharony:1999ti}
\bibitem{Aharony:1999ti}
O.~Aharony, S.~S.~Gubser, J.~M.~Maldacena, H.~Ooguri and Y.~Oz,
``Large N field theories, string theory and gravity,''
Phys.\ Rept.\  {\bf 323} (2000) 183
[arXiv:hep-th/9905111].
%%CITATION = HEP-TH 9905111;%%


%\cite{D'Hoker:2002aw}
\bibitem{D'Hoker:2002aw}
E.~D'Hoker and D.~Z.~Freedman,
``Supersymmetric gauge theories and the AdS/CFT correspondence,''
arXiv:hep-th/0201253.
%%CITATION = HEP-TH 0201253;%%


%\cite{Mikhailov:2002bp}
\bibitem{Mikhailov:2002bp}
A.~Mikhailov,
``Notes on higher spin symmetries,''
arXiv:hep-th/0201019.
%%CITATION = HEP-TH 0201019;%%

%\cite{Fradkin:ks}
\bibitem{Fradkin:ks}
E.~S.~Fradkin and M.~A.~Vasiliev,y
``On The Gravitational Interaction Of Massless Higher Spin Fields,''
Phys.\ Lett.\ B {\bf 189} (1987) 89.
%%CITATION = PHLTA,B189,89;%%

%\cite{Fradkin:1986qy}
\bibitem{Fradkin:1986qy}
E.~S.~Fradkin and M.~A.~Vasiliev,
``Cubic Interaction In Extended Theories Of Massless Higher Spin Fields,''
Nucl.\ Phys.\ B {\bf 291} (1987) 141.
%%CITATION = NUPHA,B291,141;%%

%\cite{Vasiliev:1995dn}
\bibitem{Vasiliev:1995dn}
M.~A.~Vasiliev,
``Higher-spin gauge theories in four, three and two dimensions,''
Int.\ J.\ Mod.\ Phys.\ D {\bf 5} (1996) 763
[arXiv:hep-th/9611024].
%%CITATION = HEP-TH 9611024;%%

%\cite{Konstein:2000bi}
\bibitem{Konstein:2000bi}
S.~E.~Konstein, M.~A.~Vasiliev and V.~N.~Zaikin,
``Conformal higher spin currents in any dimension and AdS/CFT  correspondence,''
JHEP {\bf 0012} (2000) 018
[arXiv:hep-th/0010239].
%%CITATION = HEP-TH 0010239;%%

%\cite{Klebanov:2002ja}
\bibitem{Klebanov:2002ja}
I.~R.~Klebanov and A.~M.~Polyakov,
``AdS dual of the critical O(N) vector model,''
arXiv:hep-th/0210114.
%%CITATION = HEP-TH 0210114;%%

%\cite{Lang:1992pp}
\bibitem{Lang:1992pp}
K.~Lang and W.~R\"uhl,
``The Critical O(N) sigma model at dimensions 2 $<$ d $<$ 4: A list of quasiprimary fields,''
Nucl.\ Phys.\ B {\bf 402} (1993) 573.
%%CITATION = NUPHA,B402,573;%%

%%\cite{Lang:1991kp}
%\bibitem{Lang:1991kp}
%K.~Lang and W.~Ruhl,
%``The Critical O(N) sigma model at dimension 2 < d < 4 and order 1/n**2: Opera%tor product expansions and renormalization,''
%Nucl.\ Phys.\ B {\bf 377} (1992) 371.
%%CITATION = NUPHA,B377,371;%%

\bibitem{LR}
K.~Lang and W.~R\"uhl,
``The Critical O(N) sigma model at dimensions $2<d<4$: Fusion coefficients and anomalous dimensions,''
Nucl.\ Phys.\ B {\bf 400} (1993) 597.
%%CITATION = NUPHA,B400,597;%%

%\cite{Fradkin:1990dk}
\bibitem{Fradkin:1990dk}
E.~S.~Fradkin and M.~Y.~Palchik,
``Exactly Solvable Conformally Invariant Quantum Field Models In D-Dimensions,''
Int.\ J.\ Mod.\ Phys.\ A {\bf 5} (1990) 3463.
%%CITATION = IMPAE,A5,3463;%%

%\cite{Vasilev:yc}
\bibitem{Vasilev:yc}
A.~N.~Vasilev, Y.~M.~Pismak and Y.~R.~Khonkonen,
``Simple Method Of Calculating The Critical Indices In The 1/N Expansion,''
Theor.\ Math.\ Phys.\  {\bf 46} (1981) 104
[Teor.\ Mat.\ Fiz.\  {\bf 46} (1981) 157].
%%CITATION = TMPHA,46,104;%%

%\cite{Vasilev:dg}
\bibitem{Vasilev:dg}
A.~N.~Vasilev, Y.~M.~Pismak and Y.~R.~Khonkonen,
``1/N Expansion: Calculation Of The Exponents Eta And Nu In The Order 1/N**2 For Arbitrary Number Of Dimensions,''
Theor.\ Math.\ Phys.\  {\bf 47} (1981) 465
[Teor.\ Mat.\ Fiz.\  {\bf 47} (1981) 291].
%%CITATION = TMPHA,47,465;%%

%\cite{Vasilev:dc}
\bibitem{Vasilev:dc}
A.~N.~Vasilev, Y.~M.~Pismak and Y.~R.~Khonkonen,
``1/N Expansion: Calculation Of The Exponent Eta In The Order 1/N**3 By The Conformal Bootstrap Method,''
Theor.\ Math.\ Phys.\  {\bf 50} (1982) 127
[Teor.\ Mat.\ Fiz.\  {\bf 50} (1982) 195].
%%CITATION = TMPHA,50,127;%%

%\cite{Leonhardt:2002ta}
\bibitem{Leonhardt:2002ta}
T.~Leonhardt and W.~R\"uhl,
``General graviton exchange graph for four point functions in the AdS/CFT  correspondence,''
arXiv:hep-th/0210195.
%%CITATION = HEP-TH 0210195;%%




\end{thebibliography}
\end{document}